\documentclass[11pt]{article}

\usepackage{graphicx} 
\usepackage{psfrag}
\usepackage{amsmath,amssymb}
\usepackage{amsthm}
\usepackage{verbatim}
\usepackage[usenames,dvipsnames]{color}
\usepackage{epstopdf}
\usepackage{url}
\usepackage{hyperref}

\usepackage{graphicx}
\usepackage{epstopdf}
\usepackage{color}

\usepackage{amsmath,amsfonts}
\newcommand{\bqn}{\begin{eqnarray}}
\newcommand{\eqn}{\end{eqnarray}}
\newcommand{\bq}{\begin{eqnarray*}}
\newcommand{\eq}{\end{eqnarray*}}

\newcommand{\red}[1]{\textcolor{red}{#1}}

\begin{document}

\title{Kernel Regression on Manifolds and Its Application\\ to Modeling
Disconnected Anatomic Structures}
 
\author{Moo K. Chung$^{1,2,3}$, 
 Nagesh Adluru$^{3}$, Houri K. Vorperian$^{2}$\\
$^1$Department of Biostatistics and Medical Informatics\\
$^2$Vocal Tract Laboratory, Waisman Center\\
$^3$Waisman Laboratory for Brain Imaging and Behavior\\
University of Wisconsin-Madison, USA
\\
\vspace{0.3cm}
\red{\tt mkchung@wisc.edu}
}

\maketitle

\pagenumbering{arabic}

\begin{abstract}
We present a unified statistical approach to modeling disconnected 3D anatomical structures extracted from medical images. 
Due to image acquisition and preprocessing noises, it is expected the imaging data is noisy. 
The surface coordinates of the structures are regressed using the weighted linear combination of Laplace-Beltrami (LB) eigenfunctions to smooth out noisy data and perform statistical analysis.
The method is applied in characterizing the 3D growth pattern of human hyoid bone between ages 0 and 20. We detected a significant age effect on localized parts of the hyoid bone.
\end{abstract}


\section{Introduction}
For normally developing children, age and gender could be major factors that affect the functions and structures of growing hyoid bone. As in other developmental studies \cite{chung.2003.ni,vorperian.2011}, we expect highly localized complex growth pattern to emerge between ages 0 and 20 in the hyoid bone. It is expected the growth to be outward with respect to the surface of the bone. However, it is unclear what specific parts of the hyoid bone are growing. This provides a biological motivation for a need to develop a local surface-based morphometric technique beyond simple volumetric techniques that cannot detect localized subtle anatomical changes along the hyoid bone surface. 

The end results of existing surface-based morphometric studies in medical imaging are statistical parametric maps (SPM) that shows statistical significance of growth at each surface mesh vertex \cite{chung.2003.ni,qiu.2008,xu.2008}. In order to obtain stable and robust SPM, various signal smoothing and filtering methods have been proposed. Among them, diffusion equations, kernel smoothing, and wavelet-based approaches are probably most popular. Diffusion equations have been widely used in image processing as a form of noise reduction starting with Perona and Malik in 1990's \cite{perona.1990}. Although numerous techniques have been developed for performing diffusion along surfaces
\cite{chung.2003.cvpr,andrade.2001,tang.1999,sochen.1998,malladi.2002,taubin.2000},  
most  approaches are nonparametric and requires finite element or finite difference schemes which are known to suffer various numerical instabilities \cite{chung.2005.IPMI}. 

Kernel smoothing based models have been also proposed for surface and manifolds data \cite{belkin.2006,chung.2005.IPMI}. The kernel methods basically smooth data as weighted average of neighboring mesh vertices using mostly a Gaussian kernel and its iterative application is supposed to approximates the diffusion process. Recently, wavelets have been popularized for surface and graph data. Spherical wavelets have been used on brain surface data that has been mapped onto a sphere  \cite{nain.2007,bernal.2008}.
Since wavelet basis functions have local supports in both space and scale, the
wavelet coefficients from the scale-space decomposition using the spherical wavelets provides shape features that describe local shape variation at a variety of scales and spatial locations. 
However, spherical wavelets have an intrinsic problem that they require to establish a smooth mapping from the surface to a unit sphere, which introduces a serious metric distortion. 
The spherical mapping such as conformal mapping introduces serious metric distortion which usually compounds SPM. Furthermore, such basis functions defined on sphere seem to be suboptimal rather than those directly defined on anatomical surface, in detecting locations or scales of shape variations.  
To remedy the limitation of spherical wavelets, spectral graph wavelet transform defined on a graph has been applied to arbitrary surface meshes by treating surface meshes as graphs \cite{antoine.2010,hammond.2011,kim.2012.NIPS}. Wavelet transform is a powerful tool decomposing a signal or function into a collection of components localized at both location and scale. 
Although all three methods (diffusion-, kernel- and wavelet-based) look different from each other, it is possible to develop a unified framework that relates all of them in a coherent mathematical framework.  

Starting with a symmetric positive definite kernel, we propose a unified kernel regression framework within the Hilbert space theory. The proposed kernel regression works for any symmetric positive definite kernel, which behaves like weights between two functional data. We show how this facilitates a coherent statistical inference for functional signals defined on an arbitrary manifold. The focus of the paper is on the development of the proposed kernel regression on manifolds.
The outline of this paper is as follows.
\begin{itemize}
\item[(i)] First, we present a unified bivariate kernel regression that is related to diffusion-like equations on manifolds. The proposed kernel regression inherits various mathematical and statistical properties of diffusion-like equations.

\item[(ii)] We establish the relationship between the kernel regression and recently popular spectral graph wavelets for manifolds. The proposed kernel regression is shown to be equivalent to the wavelet transform. This mathematical equivalence levitates a need for constructing wavelets using a complicated computational machinery as often done in previous diffusion wavelet constructions  \cite{antoine.2010,hammond.2011,kim.2012.NIPS}. 

\item[(iii)] A unified statistical inference framework is developed for a CT imaging application by linking the kernel regression to the random field theory \cite{taylor.2007,worsley.2004}. This levitates the need for using time consuming nonparametric procedures such as false discovery rates (FDR) or permutation tests that do not have explicate control over the scale and smoothness of models. 

\item[(iv)] Finally, we illustrate how the kernel regression procedure can be used to localize the disconnected hyoid bone growth pattern in human. 
\end{itemize}

\begin{figure}[t]
\centering
\includegraphics[width=1\linewidth]{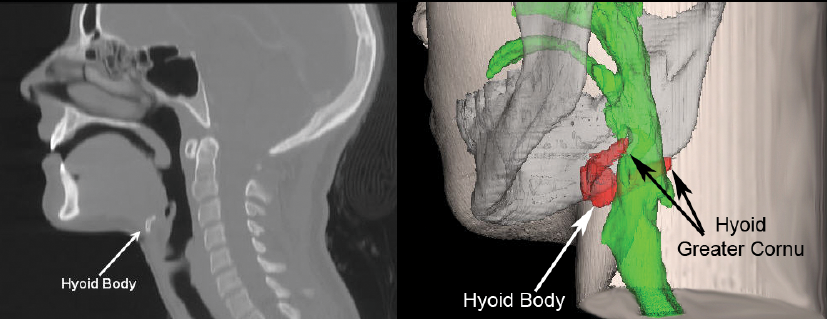}
\caption{CT image showing the location of the hyoid bone and 3D model showing the relative location of the hyoid bone with respect to the mandible (gray) and vocal tract structures (green).}
\label{fig:hyoidbone}
\end{figure}

\section{Preliminary}
First, let us illustrate two statistical problems in an Euclidean space that motivate the development of the proposed kernel regression in manifolds.

Consider measurements $f_i$ sampled at $p_i \in \mathbb{R}^d$. The measurements are usually modeled as  
$$f_i = h(p_i) + \epsilon_i$$
with mean zero noise $\epsilon_i$ and unknown mean function $h$ that has to be estimated. In the traditional  kernel regression framework \cite{belkin.2006,fan.1996,oztireli.2009}, the mean function $h$ is estimated in the weighted least squares fashion: 
$$\widehat{h}(p) = \sum_{j=1}^k G(p- p_i)f_i,$$ 
where $G$ is a given Nadaraya-Waton type of normalized kernel. In the local polynomial regression framework, $h$ is estimated as
\bqn \widehat{h}(p) =  \arg \min_{\beta_0 \cdots \beta_k} \sum_{i=1}^n G(p-p_i) \Big| f_i - \sum_{j=0}^k \beta_j (p-p_i)^j \Big|^2. \label{eq:localpolynomial2}\eqn
Often normalized Gaussian kernels are used for $G$. 
In many related local polynomial or kernel regression frameworks, kernel $G$ and polynomial basis $p^j$ are translated by the amount of $p_i$ in fitting the data locally. In this fashion, at each data point $p_i$, exactly the same shape of kernel and distance are used. However, one immediately encounters a difficulty of directly generalizing the Euclidean formulation (\ref{eq:localpolynomial2}) to an arbitrary surface since it is unclear how to translate the kernel and basis in a coherent fashion. To remedy this problem, many recent kernel regression framework on manifolds use bivariate kernel $G(p,q)$ and bypass the problem of translating a univariate kernel \cite{belkin.2006}. By simply changing the second argument, it has the effect of translating the kernel.

A similar problem is also encountered in wavelets in a Euclidean space. Consider a wavelet basis $W_{t,q}(p)$ obtained from a mother wavelet $W$ with scale and translation parameters $t$ and $q$:
\bqn W_{t, q}(p) = \frac{1}{t}W \big(\frac{p-q}{t} \big). \label{eq:Wtq} \eqn
Scaling a function on a surface is trivial. But the difficulty arises when one tries to define a mother wavelet and translate it on a surface. It is not straightforward to generalize the Euclidean formulation (\ref{eq:Wtq}) to an arbitrary manifold. 
If one tries to modify the existing spherical wavelets to an arbitrary surface \cite{nain.2007,bernal.2008}, one also encounters the lack of regular grids on the surface. The recent work based on the spectral graph wavelet transform bypasses this problem also by taking bivariate kernel as a mother wavelet \cite{antoine.2010,hammond.2011,mahadevan.2006, kim.2012.NIPS}.
To remedy these two different but related problems, we propose to use a bivariate kernel and bypass the problem of translating a univariate kernel. By simply changing the second argument, it has the effect of translating the kernel. 

\section{Methods}

In many anatomical studies in 
medical imaging, measurements are sampled densely at each voxel, 
so it is more practical to model the measurements as a function. Consider a functional measurement $f$ defined on a manifold $\mathcal{M} \subset \mathbb{R}^d$. We assume the following additive model:
\bqn f(p) =  h(p) + \epsilon(p), \label{eq:model1} \eqn
where $h$ is the unknown signal and $\epsilon$ is a zero-mean random field, possibly Gaussian. The manifold $\mathcal{M}$ can be a single connected or multiple disjoint components as our hyoid bone application. We further assume $f \in L^2(\mathcal{M})$, the space of square integrable functions on $\mathcal{M}$ with the inner product 
$$\langle f,g \rangle = \int _{\mathcal{M}} f(p)g(p) \;d\mu(p),$$ 
where $\mu$ is the Lebesgue measure. 
Define a self-adjoint operator $\mathcal{L}$ satisfying
$$\langle g_1, \mathcal{L} g_2 \rangle = \langle
\mathcal{L}g_1, g_2 \rangle$$ 
for all $g_1, g_2 \in
L^2(\mathcal{M})$. Then $\mathcal{L}$ induces the eigenvalues $\lambda_j$ and
eigenfunctions $\psi_j$ on $\mathcal{M}$:
\bqn \mathcal{L} \psi_j = \lambda_j \psi_j. \label{eq:eigen}\eqn 
Without loss of generality, we can order the
eigenvalues $0 = \lambda_0 \leq \lambda_1 \leq \lambda_2 \leq
\cdots.$
The eigenfunctions $\psi_j$ form an orthonormal basis in $L^2(\mathcal{M})$. We will consider 
a smooth symmetric positive definite kernel of the form
\bqn K(p,q) = \sum_{j=0}^{\infty} \tau_j \psi_{j}(p)\psi_{j}(q) \label{eq:Kpq11}\eqn
for some $\tau_j$ in this paper. The constants $\tau_j$ are identified as follows. Apply the kernel convolution on the eigenfunction $\psi_j$:
\bqn K*\psi_j(p) =  \int_{\mathcal{M}} K(p,q) \psi_j(q)\; d\mu(q). 
\label{eq:Kf1} 
\eqn
Substituting (\ref{eq:Kpq11}) into (\ref{eq:Kf1}), we have
$K*\psi_j(p) = \tau_j \psi_j(p)$
indicating $\tau_j$  and $\psi_j$ must be the eigenvalues and eigenfunctions of the convolution (\ref{eq:Kf1}).

\noindent{\bf Example 1.}  For $\tau_j = e^{-\lambda t}$, we have heat kernel 
\bqn K(p,q) = \sum_{j=0}^{\infty} e^{-\lambda t} \psi_{j}(p)\psi_{j}(q) \label{eq:Kpq11}\eqn
that has been often discussed in numerous studies but without much theoretical justification \cite{hendriks.1990,seo.2010.MICCAI,kim.2011.PSIVT,chung.2015.MIA}. For this study, we will denote the heat kernel as $H_t(p,q)$ to explicitly show that the spread of the kernel is determined by $t$, diffusion time.

\begin{figure}[t]
\centering
\includegraphics[width=0.7\linewidth]{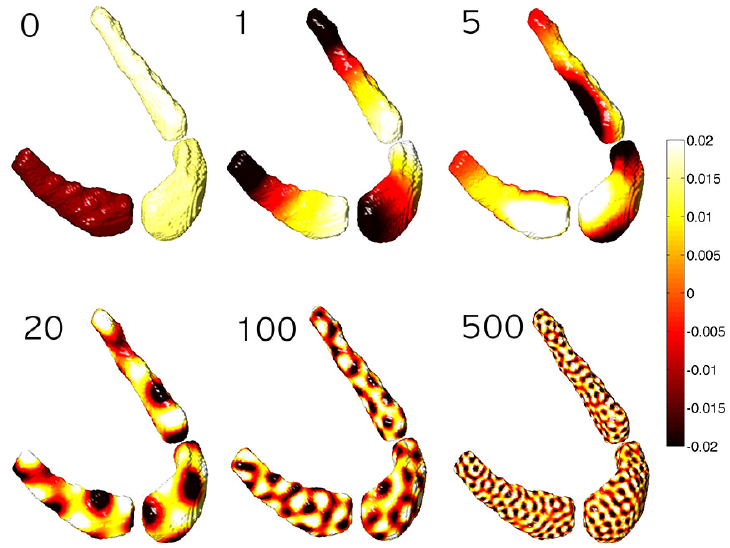}  
\caption{Laplace-Beltrami eigenfunctions  $\psi_j$ of various degrees $(j=0, 1, 5, 20, 100, 500)$ on the template. The first eigenfunction is constant in each component. As the degree increases, the spatial frequency increases.}
\label{fig:2_md_eigfs}
\end{figure}

\begin{figure}[t]
\centering
\includegraphics[width=0.7\linewidth]{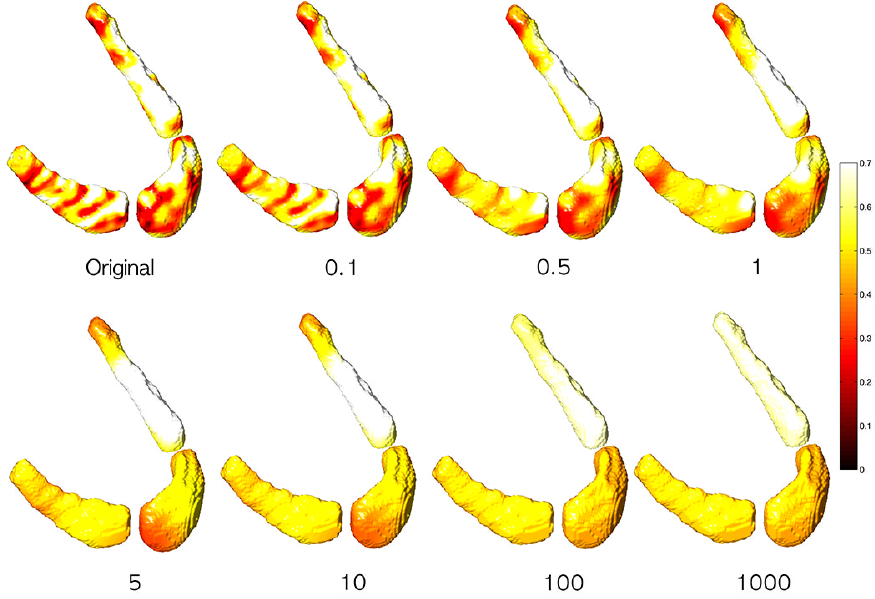}
\caption{\footnotesize  Heat kernel regression with different bandwidth between 0.1 and 1000. As the bandwidth increases, the kernel regression becomes inversely proportional to the square root of the surface area.}
\label{fig:wavelets}
\end{figure}

\subsection{Kernel regression on manifolds}
Consider subspace $\mathcal{H}_k \subset L^2(\mathcal{M})$ spanned by the orthonormal basis $\{\psi_j \}$, i.e. $$\mathcal{H}_k =\{ \sum_{j=0}^k \beta_j \psi_j(p): \beta_j \in
\mathbb{R}\}.$$
Then the least squares  estimation (LSE) of $h$ in $\mathcal{H}_k$ is given by the shortest distance from $f$ to $\mathcal{H}_k$:
\bqn \widehat{h}(p) = 
 \arg\min_{h \in \mathcal{H}_k} \int_{\mathcal{M}} \big| f(p) - h(p) \big|^2 \; d\mu(p) = \sum_{j=0}^k f_j \psi_j(p), \label{eq:LSE}\eqn
where $f_j = \langle f,\psi_j \rangle$ are the Fourier coefficients. Figure \ref{fig:wavelets} shows an example of LSE with $\mathcal{L}$ as the Laplace-Beltrami operator and $k=1000$. This is the usual Fourier series expansion that tends to suffer  the Gibbs phenomenon, i.e., ringing artifact \cite{chung.2007.TMI,gelb.1997} for compact surfaces (Example 1).

The Gibbs phenomenon can be effectively removed if the Fourier series expansion converges fast enough as the number of basis functions goes to infinity. By weighting the Fourier coefficients exponentially smaller, we can make the representation converges faster; this can be achieved by additionally weighting the squared residuals in equation (\ref{eq:LSE}) with some weights. Thus, we propose to estimate $h$ by minimizing the weighted distance to the space $\mathcal{H}_k$:
\bqn \widehat{h}(p) = \arg\min_{h \in \mathcal{H}_k} \int_{\mathcal{M}}\int_{\mathcal{M}} K(p,q) \Big| f(q) - h(p) \Big|^2 \; d\mu(q)\; d\mu(p). \label{eq:HKregression}\eqn
Without loss of generality, we will assume the kernel to be a probability distribution so that
$$\int_{\mathcal{M}} K(p,q) \; d\mu(q) = 1$$
for all $p \in \mathcal{M}$. The solution of (\ref{eq:HKregression}) has the following analytic expression:\\

\noindent{\bf Theorem 1.} 
\bq \widehat{h}(p)  = \arg\min_{h \in \mathcal{H}_k}\int_{\mathcal{M}}\int_{\mathcal{M}} K(p,q) \Big| f(q) - h(p) \Big|^2 \; d\mu(q)\; d\mu(p) = \sum_{j=0}^k  \tau_j f_j \psi_j,
\eq
\label{theorem:kernel1}
where $f_j = \langle f, \psi_j \rangle$ are Fourier coefficients.

\noindent{\em Proof.} 
Any function $h \in \mathcal{H}_k$ can be expressed as 
\bqn h(p) = \sum_{j=0}^k \beta_j \psi_j(p) \label{eq:hp}.\eqn
Then by plugging (\ref{eq:hp}) into the inner integral $I(p)$, 
it becomes
$$I(p) =\int_{\mathcal{M}} K(p,q)\Big| f(q) - \sum_{j=0}^k  \beta_j \psi (p) \Big|^2\;d\mu(q).$$
Simplifying the expression, we obtain 
\bqn I(p) = \sum_{j=0}^k \sum_{j'=0}^k  \psi_j (p)\psi_{j'}(p)\beta_j \beta_{j'} - 2K* f(p)\sum_{j=0}^k  \psi_j(p)\beta_{j} + K*f^2(p).\label{eq:QP}\eqn
The kernel can be written as
\bqn K(p,q) = \sum_{j'=0}^{\infty} \tau_{j'} \psi_{j'}(p)\psi_{j'}(q). \label{eq:Kpq}\eqn
The convolution 
is then written as
$$K*f(p) = \sum_{j'=0}^{\infty} \tau_{j'} f_{j'} \psi_{j'}(p).$$

Since $I$ is an unconstrained positive semidefinite quadratic program (QP) in $\beta_j$, there is no unique global minimizer of $I$ without additional linear constraints. Integrating $I$ further with respect to $d\mu(p)$, we collapses (\ref{eq:QP})  to a positive definite QP, which yields a unique global minimizer:
$$\int_{\mathcal{M}} I(p) \;d\mu(p) = \sum_{j=0}^k   \beta_{j}^2 - 2\sum_{j=0}^k \tau_j f_j \beta_{j} + \mbox{ const}.$$
The minimum of the above integral is obtained when all the partial derivatives with respect to $\beta_j$ vanish, i.e.
$$\int_{\mathcal{M}} \frac{\partial I}{\partial \beta_j} \;d\mu(p) = 
2\beta_j - 2\tau_j  f_j  =0$$
for all $j$. Hence $\sum_{j=0}^k  \tau_j f_j \psi_j$ must be the unique minimizer. \qed

Theorem 1 generalizes the weighted spherical harmonic (SPHARM) representation on a unit sphere to an arbitrary manifold \cite{chung.2008.sinica}. Theorem 1 implies that the kernel regression can be performed by simply computing the Fourier coefficients $f_j = \langle f, \psi_j \rangle$ without doing any numerical optimization. The numerically difficult optimization problem is reduced to the problem of computing Fourier coefficients. 
If the kernel $K$ is a Dirac-delta function, the kernel regression simply collapses to the least squares estimation (LSE) which results in the standard Fourier series, i.e.
$$\widehat h(p) = \arg\min_{h \in \mathcal{H}_k}\int_{\mathcal{M}} \Big| f(q) - h(q) \Big|^2 \; d\mu(q)  = \sum_{j=0}^k  f_j \psi_j.$$
It can be also shown that as $k \to \infty$, the kernel regression $$\widehat{h}= \sum_{j=0}^k \tau_j f_j \psi_j$$ converges to convolution $K*f$ establishing the connection to the manifold-based kernel smoothing framework \cite{belkin.2002,chung.2005.IPMI}. Hence, asymptotically the proposed kernel regression should inherit many statistical properties of kernel smoothing. 

\subsection{Properties of kernel regression}
The kernel regression can be shown to be related to the following diffusion-like Cauchy problem. \\

\noindent{\bf Theorem 2.}
\label{theorem:cauchy} For an arbitrary self-adjoint differential operator $\mathcal{L}$, the unique solution of the following initial value problem
\bqn \frac{\partial g(p,t)}{\partial t} + \mathcal{L} g(p,t) =0 \label{eq:cauchy1}, g(p,t=0) =f(p) \label{eq:cauchy2}
\eqn
is given by
\bqn g(p,t) = \sum_{j=0}^{\infty} e^{-\lambda_j t} f_j \psi_j(p).\label{eq:cauchy.solution}\eqn

\noindent{\em Proof.} 
For each fixed $t$,
$g(p,t)$ can be written as
\bqn g(p,t) = \sum_{j=0}^{\infty} c_j(t) \psi_j(p)
\label{eq:expansion}.\eqn 
Then 
\bqn \mathcal{L} g(p,t) = \sum_{j=0}^{\infty} c_j(t) \lambda_j \psi_j(p)
\label{eq:expansion2}.\eqn 
Substituting (\ref{eq:expansion}) and (\ref{eq:expansion2})
into (\ref{eq:cauchy1}), we obtain 
\bqn \frac{\partial c_j(t)}{\partial t} + 
\lambda_j c_j(t)=0 \label{eq:ODE}\eqn 
for all $j$. The solution of equation
(\ref{eq:ODE}) is given by $c_j(t) = b_je^{-\lambda_j t}$. So we
have a solution $$g(p,t) = \sum_{j=0}^{\infty} b_j e^{-\lambda_j t}
\psi_j(p).$$ 
At $t=0$, we have
$$g(p,0) = \sum_{j=0}^{\infty} b_j \psi_j(p) = f(p).$$
The coefficients $b_j$ must be the Fourier coefficients, i.e.
$$ b_j =  \langle f,\psi_j \rangle = f_j.$$
\qed

For a particular choice of kernel $K$ with $\tau_j = e^{-\lambda_j t}$, the proposed kernel regression $\widehat{h}= \sum_{j=0}^k \tau_j f_j \psi_j$ should converge to the solution of the diffusion-like equation.

\noindent{\bf Example 2.} If $\mathcal{L}$ is the Laplace-Beltrami operator, (\ref{eq:cauchy2}) becomes an isotropic diffusion equation as a special case and we are then dealing with heat kernel 
$$H_t(p,q) = \sum_{j=0}^{\infty} e^{-\lambda_j t} \psi_j(p)\psi_j(q),$$
which is often explored mathematical objects in various fields \cite{belkin.2002,chung.2005.IPMI}.

In order to construct wavelets on an arbitrary graph and mesh, diffusion wavelet transform has been proposed recently \cite{antoine.2010,hammond.2011,kim.2012.NIPS}. The diffusion wavelet construction has been fairly involving so far.  However, it can be shown to be a special case of the proposed kernel regression and the proposed method is substantially simpler to construct. Following the notations in \cite{antoine.2010,hammond.2011,kim.2012.NIPS},  
 diffusion wavelet $W_{t,p}(p)$ at position $p$ and scale $t$ is given by
$$W_{t,q}(p) = \sum_{j=0}^k g(\lambda_j t)\psi_j(p)\psi_j(q),$$
for some scale function $g$. If we let $\tau_j = g(\lambda_j t)$, the diffusion wavelet transform  
is given by 
$$\langle W_{t,p}, f \rangle = \int_{\mathcal{M}} W_{t,q}(p)f(p) \;d\mu(p) = \sum_{j=0}^k \tau_j f_j \psi_j(q),$$ 
which is the exactly kernel regression we introduced. Hence, the diffusion wavelet transform can be simply obtained by doing the kernel regression without an additional wavelet machinery \cite{kim.2012.NIPS}. 
Further, if we let $g(\lambda_j t) = e^{-\lambda_j t}$, we have 
$$W_{t,p}(q) = H_t(p,q),$$
which is a heat kernel. The bandwidth $t$ of heat kernel controls resolution while the translation is done by shifting one argument in the kernel.

\begin{figure}[t]
 \centering
\includegraphics[width=0.7\linewidth]{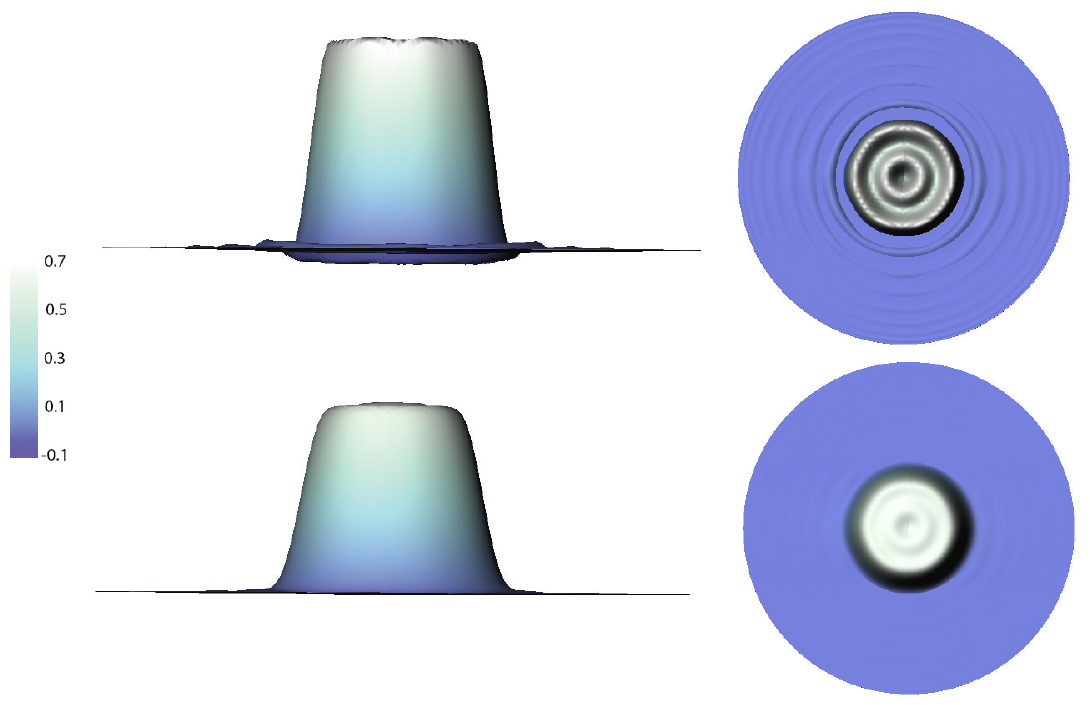}
  \caption{\footnotesize The Gibbs phenomenon on a hat shaped simulated surface showing the ringing effect on the traditional Fourier series expansion (top) and the reduced effect on the heat kernel regression (bottom). 7225 basis functions were used for the both cases and the bandwidth $t = 0.001$ is used for the kernel regression.}
\label{fig:midus-gibbs}
\end{figure}

Although the kernel regression is constructed using global basis functions $\psi_j$, the kernel regression at each point $p$ coincides with the diffusion wavelet transform at that point. Hence, just like wavelets, the kernel regression will have the localization property of wavelets. This is demonstrated in the following example:

\noindent{\bf Example 3.} A hat-shaped step function is simulated in 3D as $z=1$ for $x^2 + y^2 < 1$ and $z=0$ for $1 \leq  x^2 + y^2 \leq 2$ (Figure \ref{fig:midus-gibbs}).  Then the step function is reconstructed using the Fourier series expansion via LSE (top) and kernel regression (bottom). In the both cases,  up to 7225 basis functions were used. For the kernel regression, the heat kernel with bandwidth $t=0.0001$ is used. LSE clearly shows the visible Gibbs phenomenon, i.e., ringing artifact \cite{chung.2007.TMI,gelb.1997} compared to the kernel regression.

\subsection{Numerical Implementation}
The Laplace-Beltrami operator is chosen as the self-adjoint operators $\mathcal{L}$ of choice. The eigenfunctions of the Laplace-Beltrami operator on an arbitrary curved surface is analytically unknown. So it is necessary to discretize (\ref{eq:eigen}) using the Cotan formulation  as a generalized eigenvalue problem \cite{
zhang.2007,qiu.2006}:
\begin{equation}
\mathbf{C}\psi = \lambda \mathbf{A} \boldsymbol{\psi},
\label{eq:LMA}
\end{equation}
\noindent
where $\mathbf{C}$ is the stiffness matrix, $\mathbf{A}$ is the mass matrix and $\boldsymbol{\psi}=(\psi(p_1), \cdots, \psi(p_n))'$ is the eigenfunction evaluated at $n$ mesh vertices. 
Once we obtained the basis functions $\psi_j$, the corresponding Fourier coefficients $\beta_j$ are estimated as
\bq
\label{eq:smoothing-FEMbeta}
	\beta_j = {\bf f}' {\bf A} \boldsymbol{\psi}_j, 
\eq
where 
${\bf f} = (f(p_1), \cdots, f(p_n))'$
 and 
$\boldsymbol{\psi}_j= (\psi_j(p_1), \cdots, \psi_j(p_n))'$ \cite{zhang.2007}. 
Figure \ref{fig:2_md_eigfs} shows few representative LB-eigenfunctions on the hyoid surface. For  heat kernel regression, we used the bandwidth $\sigma =5$ and $500$ LB-eigenfunctions on the final template. 
The number of eigenfunctions used is more than sufficient to guarantee relative error less than $0.3\%$ in our data.

\subsection{Statistical Inference}

We are interested in determining the significance of functional signals on a manifold \ref{fig:agegrouping}. 
We borrow the statistical parametric mapping (SPM) framework for analyzing and visualizing statistical tests performed on the template surface that is often used in brain image analysis \cite{andrade.2001,lerch.2005.ni,wang.2010,worsley.1995,yushkevich.2008}. Since test statistics are constructed over all mesh vertices on the surface, multiple comparisons need to be accounted. For continuous functional data,  the random field theory \cite{taylor.2007,worsley.1995,worsley.2004} is natural to use. The random field theory assumes the measurements to be smooth Gaussian random field. Heat kernel regression will make the data more smooth and Gaussian as well as increase the signal-to-noise ratio \cite{chung.2005.ni}. 


 \begin{figure}[t]
\centering
\includegraphics[width=0.7\linewidth]{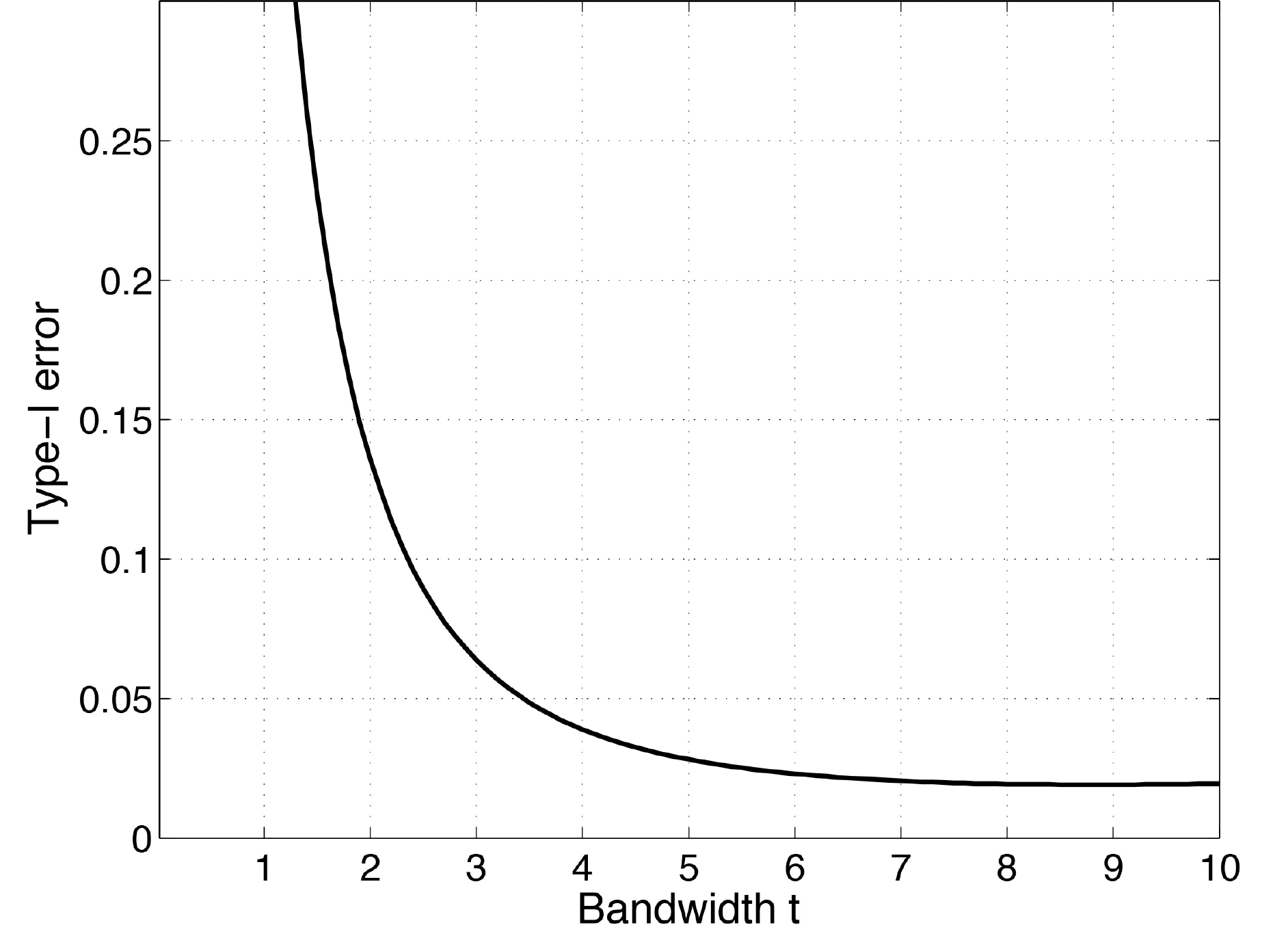}
 \caption{Type-I error plot over bandwidth $t$ of kernel regression for testing the difference between the groups I and III. As the bandwidth increases, the multiple comparisons corrected type-I error decreases. The bandwidth 5 is chosen for the study. The choice of the bandwidth around 5 does not change the over-all Type-I error much.}
\label{fig:powerplot}
\end{figure}

Consider a functional measurements $f_1, \cdots, f_n$ on manifold $\mathcal{M}$. In the simplest statistical setting, the measurements can be modeled as 
$$ f_i (p) = h(p) + \epsilon_i(p),$$
where $h$ is an unknown group level signal and $\epsilon_i$ is a zero-mean Gaussian random field \cite{worsley.2004}. At each fixed point $p$, we are assuming $\epsilon_i \sim N(0, \sigma^2)$.

We are interested in determining the significance of $h$, i.e.
\bqn H_0: h(p)=0 \mbox{ for all } p \in \mathcal{M} \; \mbox{ vs. } \;
H_1: h(p)
> 0 \mbox { for some } x \in \mathcal{M}. \label{eq:hypothesis}\eqn
Note that any point $p_0$ that gives $h(p_0) > 0$ is considered as signal. The hypothsis
(\ref{eq:hypothesis}) is an infinite dimensional multiple comparisons problem for continuously indexed hypotheses over the manifold $\mathcal{M}$. The underlying group level signal $h$ is estimated using the proposed heat kernel regression. Subsequently, a test statistic is given by a T-field $T(p)$ or a F-field, which is simply given by the square of the T-field  \cite{worsley.2004,worsley.1995}. 

For sufficiently high threshold $z$, 
the corrected type-I error of testing  hypothesis (\ref{eq:hypothesis}) is given by
$$P \Big(\sup_{p \in \mathcal{M}} T(p) > z \Big) = \sum_{j=0}^d
\mu_j(\mathcal{M})\rho_j(z), 
$$
where $\mu_d(\mathcal{M})$ is the $j$-th
Minkowski functional or intrinsic volume of $\mathcal{M}$ and
$\rho_j$ is the $j$-th Euler characteristic (EC) density of T-field. 
Since the hyoid bone is compact with no boundary but has three disconnected components, the Minkowski functionals are simply 
\bq \mu_2(\mathcal{M}) &=& \mbox{area}(\mathcal{M})/2\\
\mu_1(\mathcal{M}) &=& 0\\
\mu_0(\mathcal{M}) &=& \chi(\mathcal{M}) = 3 \times 2.
\eq 
The term $\mu_1$ is zero since there is no boundary and $\mu_0$ is simply the Euler characteristic of the template surface. Note that the Euler characteristic of a closed surface with no hole or handle is 2 and there are three such surfaces. The EC-densities of the T-field with $\nu$ degrees of freedom is given by
\bq \rho_0 (z)&=& 1 - P(T_{\nu} \leq z),\\
\rho_1 (z) & = &   \frac{1}{\sqrt{2t^2}} \cdot \frac{1}{2\pi} \Big( 1+  \frac{z^2}{\nu})^{-(\nu-1)/2}, \\
\rho_2 (z) & = & \frac{1}{2t^2} \cdot \frac{1}{(2\pi)^{3/2}} \frac{\Gamma(\frac{\nu +1}{2})}
{(\frac{\nu}{2})^{1/2} \Gamma(\frac{\nu}{2})}  z \Big( 1 + \frac{z^2}{\nu} \Big)^{-(\nu -1)/2}.    \eq
The EC-density of the F-field is similarly given in \cite{worsley.2004,
taylor.2007}. The EC-density has the kernel bandwidth $2t^2$ in the formulation so the inference is done at a particular smoothing scale. Figure \ref{fig:powerplot} shows the type-I error plot over different bandwidth $t$ of the kernel regression in our application. As the bandwidth $t$ becomes zero, the type-I error increases. When $t=0$, the kernel regression collapse to the usual Fourier series expansion. Note that the Fourier expansion with 500 LB-eigenfunctions is close to the original data without any smoothing. Hence, the proposed kernel regression can be viewed as having substantially smaller type-I error compared to the Fourier series expansion as well as  the original data demonstrating a better statistical performance. Type-II error and the statistical power can be also computed similarly. \\

\begin{figure}[t]
\begin{center}
\includegraphics[width=0.4\linewidth]{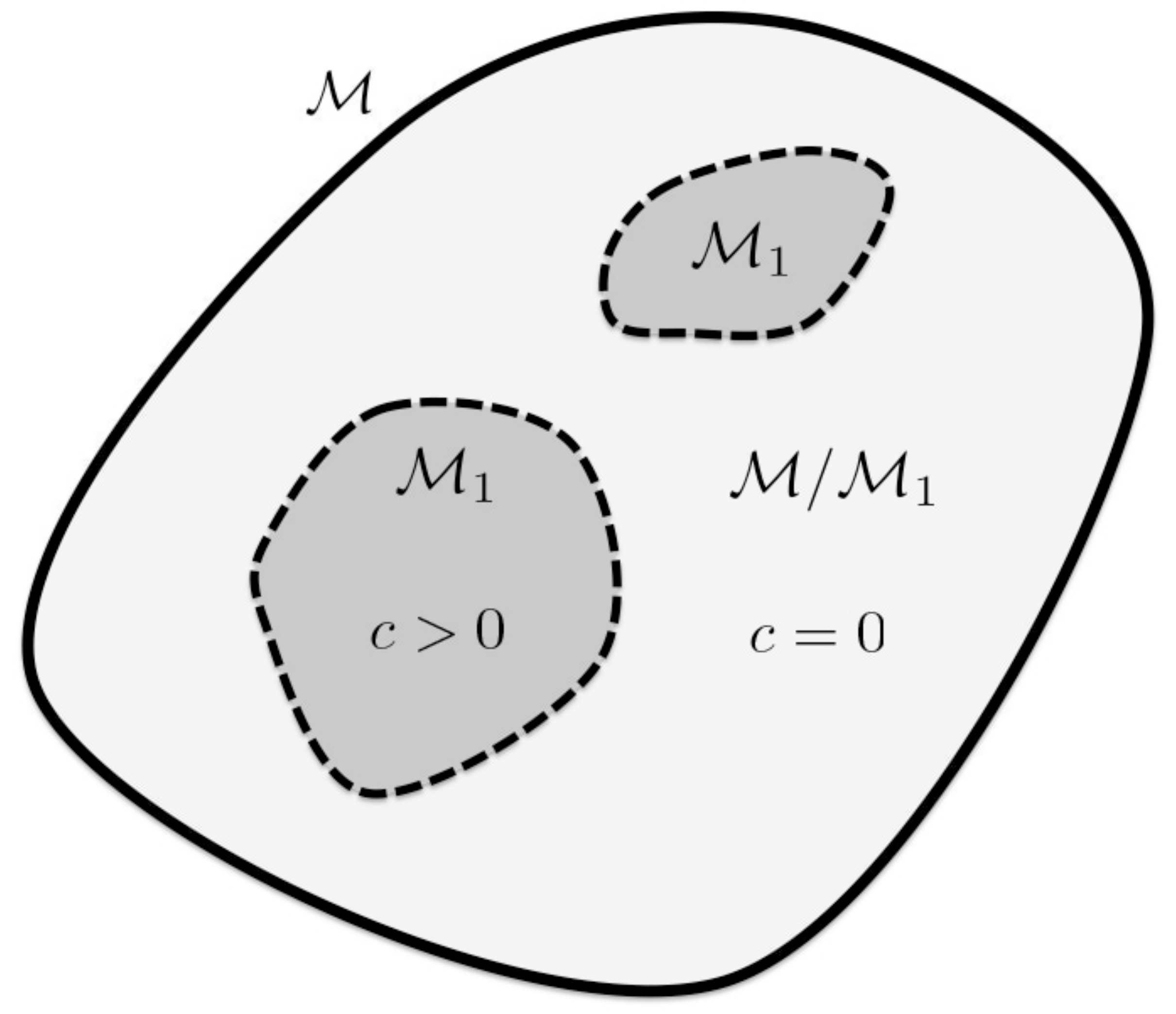}
\caption{Schematic of a case when $H_1$ is true.}
\label{fig:poweranalysisJ0}
\end{center} 
\end{figure}

\noindent{\bf Theorem 3.} The statistical power $\mathcal{P}$ of testing the hypotheses
$$ H_0: h(p)  = 0 \; \mbox{ for all } p \in \mathcal{M} \; \mbox{ vs. } \; H_1: h(p) = c\sigma > 0 \mbox{ for some } p \in \mathcal{M}. $$
using the T random field $T(p)$  is given by
$$\mathcal{P}(n) \approx 1- \exp \Big[ - \sum_{j=0}^d \mu_j(\mathcal{M}_1)\rho_j(t^*_{\alpha} - c\sqrt{n}) \Big],$$
where $t^*_{\alpha}$ is the $\alpha$-quantile given by
$$\alpha = P \Big( \sup_{p \in \mathcal{M}} T(p) > t^*_{\alpha} \Big).$$  
\noindent{\em Proof.} 
In the region $\mathcal{M}_0$ corresponding to $H_0$, 
$$f^i(p) \sim N(0, \sigma^2).$$
In the region $\mathcal{M}_1$ corresponding to $H_1$, 
$$f^i(p) \sim N( c \sigma, \sigma^2).$$
Figure \ref{fig:poweranalysisJ0} illustrates this setting.
 
Consider the test statistic
\bqn T(p) = \frac{\bar f(p)}{S(p)/\sqrt{n}}, \label{eq:T} \eqn 
where $\bar f$ and $S$ are the sample mean and standard deviation of the measurements $f^i, \cdots, f^n$. In $\mathcal{M}_0$, $T(p)$ is a T random field with $n-1$ degrees of freedom \cite{adler.1981}. In $\mathcal{M}_1$, $T(p)$ can be written as 
$$T(p) = T'(p) + \frac{c\sigma}{S(p)/\sqrt{n}},$$
where $T'(p)$ a T random field with $n-1$ degrees of freedom. Since $\sigma$ is usually estimated using the standard deviation, approximately we have $S(p) = \sigma$ and the test statistic becomes
$$T(p) = T'(p) + c\sqrt{n}.$$
At each fixed $p$, $T(p)$ is no longer a T random field but a non-central T random field \cite{hayasaka.2007}. 
Subsequently the power $\mathcal{P}$ at the given $\alpha$-level is given by
\bqn \mathcal{P}(n) &=& P \Big( \sup_{p \in \mathcal{M}_1} T(p) > t^*_{\alpha} \Big) \label{eq:Tdiff}\\
                             &=&  P \Big(  \sup_{p \in \mathcal{M}_1}  T'(p)  > t^*_{\alpha} - c\sqrt{n} \Big), \label{eq:supP}\eqn
where $t_{\alpha}^*$ is the $\alpha$-quantile of $\sup_{p \in \mathcal{M}} T(p)$ under $H_0$, i.e. 
$$\alpha = P \Big( \sup_{p \in \mathcal{M}} T(p) > t_{\alpha}^* \Big).$$

Although (\ref{eq:Tdiff}) is intractable to directly compute, we can approximate (\ref{eq:supP}) using the expected Euler characteristic (EC) method \cite{worsley.1996,worsley.2003}. The power (\ref{eq:supP}) can be written as 
$$\mathcal{P}(n) = \sum_{j=0}^d \mu_j(\mathcal{M}_1)\rho_j(t^*_{\alpha} - c\sqrt{n}),$$
where $\mu_d(\mathcal{M})$ is the $j$-th
Minkowski functional or intrinsic volume of $\mathcal{M}$ and
$\rho_j$ is the $j$-th EC-density of
T-field \cite{worsley.1998,adler.1981,taylor.2007,worsley.2003}. 
The expansion only works for sufficiently large $t^*_{\alpha} - c\sqrt{n}$. For small threshold, the power may not be bounded between 0 and 1. So it is necessary to use the exponential transform used in \cite{hayasaka.2007} to bound the power. For small $\mathcal{P}(n)$, using the Taylor expansion, we can write
$$\exp \big[ - \mathcal{P}(n) \big]  \approx 1 - \mathcal{P}(n).$$
Equivalently, it is written as 
$$\mathcal{P}(n) \approx 1 - \exp \big[ - \mathcal{P}(n) \big].$$
This transformation guarantees the power estimation to be bound between 0 and 1 \cite{hayasaka.2007}. Subsequently, the power is given by
$$\mathcal{P}(n) = 1- \exp \Big[- \sum_{j=0}^d \mu_j(\mathcal{M}_1)\rho_j(t^*_{\alpha} - c\sqrt{n}) \Big].$$
\qed

\section{Application}

\subsection{CT Imaging Data and Preprocessing}

The study consists of high resolution CT images of 70 normal subjects ages between 0 and 20 years (mean age $=$ 58.0 $\pm$ 11.3 years). 
CT scans were converted to DICOM format and Analyze $8.1$ software package (AnalyzeDirect, Inc., Overland Park, KS) was then used in segmenting binary hyoid bone images by a trained individual rater in the native space by simple image intensity thresholding and careful manual editing. A nonlinear image registration using the diffeomorphic shape and intensity averaging technique with cross-correlation as similarity metric was performed through Advanced Normalization Tools (ANTS) \cite{avants.2008}. A study-specific template was constructed. We have chosen a 12 year old subject identified as F155 as the initial template and aligned the remaining 69 hyoids to the initial template affinely to remove the overall size variability. Some subject may have larger hyoid than others so it is necessary to remove the global size differences in local shape modeling. From the affine transformed individual hyoid surfaces, we performed the diffeomorphic nonlinear image registration to the template using ANTS. 

\begin{figure}[t]
\centering
\includegraphics[width= 0.7\linewidth]{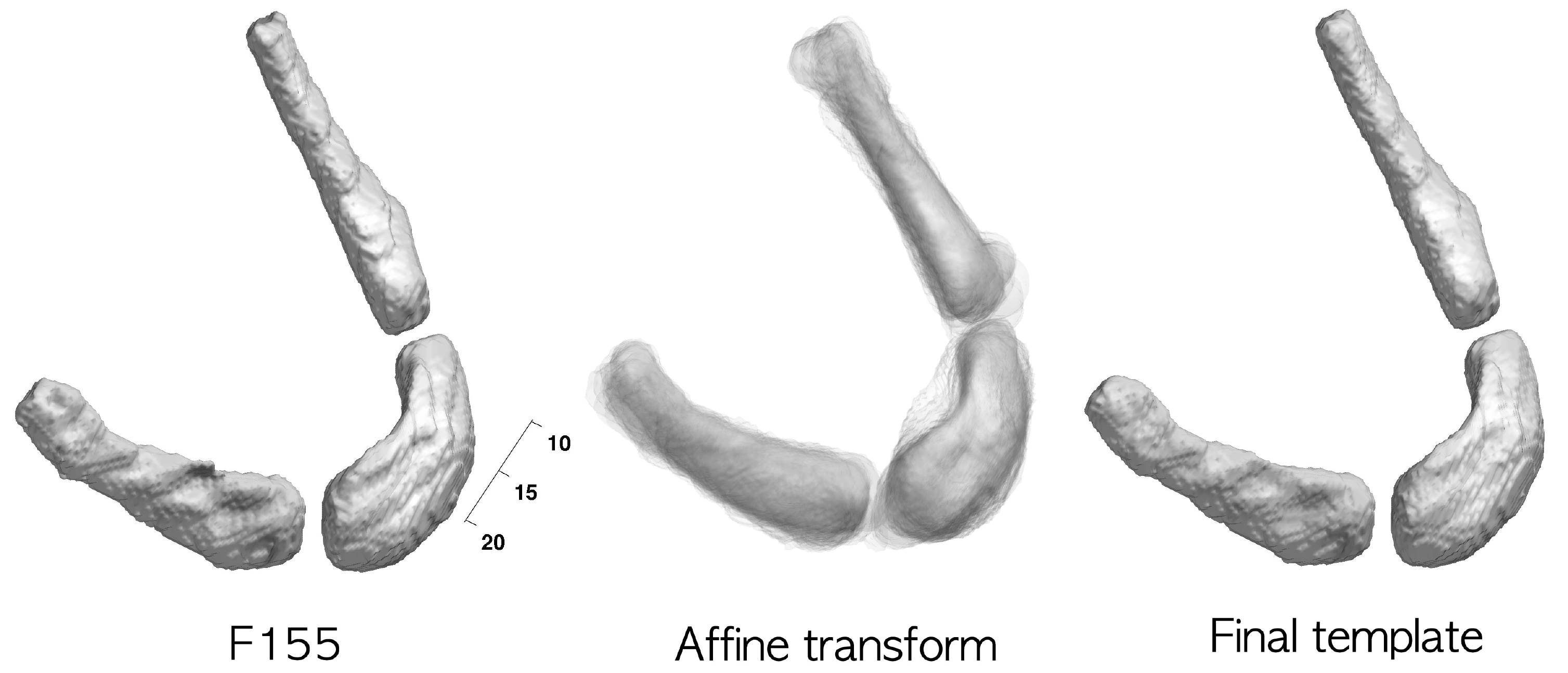} 
\caption{Left: Hyoid F155 which forms an initial template $\mathcal{M}_I$. All other mandibles are affine registered to F155. Middle: The superimposition of affine registered hyois showing local misalignments. Diffeomorphic registration is then performed to register misaligned affine transformed hyoids. 
Right: The average of deformation with respect to F155 provides the final population average template $\mathcal{M}_F$ where statistical parametric maps will be constructed.}
\label{fig:template}
\end{figure}

\begin{figure}[t]
\centering
\includegraphics[width= 1 \linewidth]{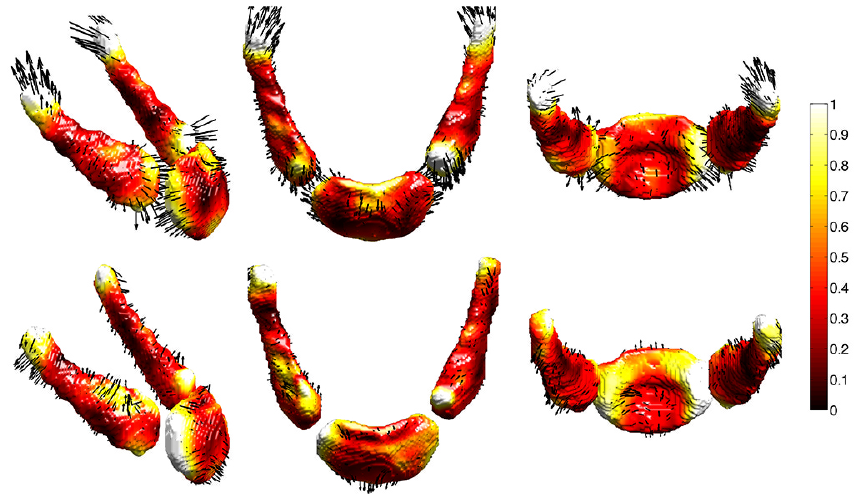} 
\caption{Hyoid bones are binned into three age groups: group I (ages 0 and 6), group II (ages 7 and 12) and group III (ages 13 and 19) and the mean displacements between the groups are visualized. Each row shows the mean group differences of the displacement: group II - group I (first row) and group III - group II (second row). The arrows are the mean displacement differences and colors indicate their lengths in mm.}
\label{fig:agegrouping}
\end{figure}

Then by averaging the inverse deformation fields from the initial template to individual hyoid, we obtain the yet another final template. Figure \ref{fig:template} shows the initial and final templates. The isosurface of the final template volume is extracted using the marching cube algorithm \cite{lorensen.1987}. Figure \ref{fig:agegrouping} shows the mean displacement differences between the groups I and II (top) and II and III (bottom). Each row shows the group differences of the displacement: group II - group I (first row) and group III - group II (second row). The arrows are the growth direction given by the mean displacement differences and colors indicate their lengths in mm. We are interested in localizing the regions of hyoid bone growth between the age groups.

70 subjects are binned into three age categories: ages between 0 and 6 years (group I), between 7 and 12 years (group II), and between 13 and 19 years (group III). There are 26, 14 and 30 subjects in group I, II and III respectively. The main biological hypothesis of interest is if there is any localized hyoid bone growth spurts between these specific age groups.

\subsection{Results}

The displacement from the template to an individual surface is obtained at each mesh vertex. Since  the length measurement provides a much easier biological interpretation, we used the length of displacement vector as a response variable among many other possible features. Since the length on the template surface is expected to be noisy due to image acquisition, segmentation and image registration errors, it is necessary perform the proposed kernel regression and subsequently reduce the type-I error and obtain more stable SPM. Figure \ref{fig:wavelets} shows an example of kernel regression on our data. The kernel regression increases the signal-to-noise ratio (SNR) and improves the smoothness and Gaussianness of data. Subsequently, the heat kernel regression of the displacement length is taken as the response variable. 
We have chosen $t=5$ as the bandwidth for the study since the bandwidth 5 is where the type-I error starts to flatten out in Figure \ref{fig:powerplot}. Note that the Fourier expansion with 500 LB-eigenfunctions is close to the original data (relative error of less than 0.3$\%$). Hence, performing the proposed kernel regression before the statistical analysis can substantially smaller type-I error demonstrating its effectiveness. 

\begin{figure}[ht]
\centering
\includegraphics[width=0.85\linewidth]{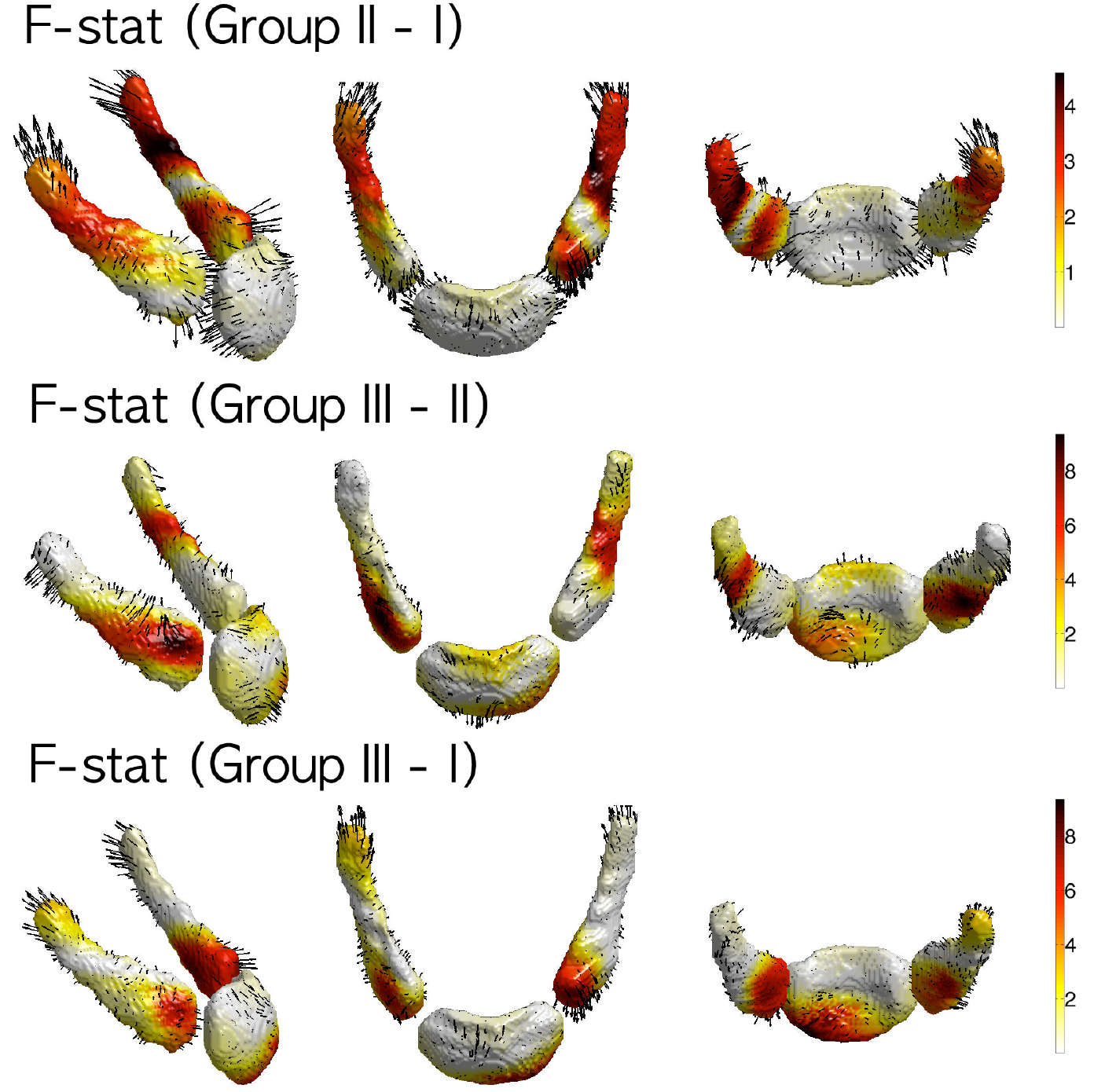}
\caption{F-statistic maps on hyoid showing age effect between the groups. The significant growth regions (red) are identified only between group II and III, and I and III. The growth is highly localized near the regions that connect the disconnected hyoid bones.}
\label{fig:surfGLM}
\end{figure}

After the displacement lengths are smoothed, we constructed the F-field, or equivalently the T-field square, for testing the length difference between the age groups I and II, II and III, and I and III showing the regions of growth spurts between different age range (Figure \ref{fig:surfGLM}). Since test statistics are constructed over all mesh vertices on the mandible, multiple comparisons were account  using the random field theory \cite{worsley.1995,worsley.2004}. 

For testing the differences between the groups I and II, II and III, and I and III, they are based on F-field with  1 and  38, 1 and 42, and 1 and  54 degrees of freedom respectively. The result is displayed in Figure \ref{fig:surfGLM}, where the significant results were only found between the groups II and III (middle), and I and III (bottom) at 0.1 level. Between the groups II and III, we obtained the maximum F-statistic value of 9.36 (right hyoid), which corresponds to the p-value of 0.041 (corrected). Between the groups I and III, we obtained the maximum F-statistic value of 10.55 (middle hyoid), which corresponds to the p-value of 0.028 (corrected). In the $F$-statistic maps for middle and bottom rows, red regions are considered as exhibiting significant growth spurts.

\section{Conclusion}
We have developed a new kernel regression framework on a manifold that unifies bivariate kernel regression, heat diffusion and wavelets in a single coherent mathematical framework. The kernel regression is both global and local in a sense it uses global basis functions to perform regression but locally equivalent to diffusion wavelet transform. The proposed framework is demonstrated to reduce type-I error in modeling shape variations compared to the usual Fourier series expansion. The method is then used in developing a statistical inference procedure for functional signals on manifolds. The whole framework 

\section*{Acknowledgment}

This work was supported by NIH Research Grants DC6282, UL1TR000427 and EB022856 and P-30 HD03352 to the Waisman Center.

\bibliographystyle{spiebib} 
\bibliography{reference.2017.08.06}

\end{document}